\documentclass[twocolumn,preprintnumbers]{revtex4}
\usepackage{graphicx}
\usepackage{dcolumn}
\usepackage{bm}
\usepackage{epsfig}
\usepackage{amsmath,amssymb}

\usepackage{CJKutf8}

\draft

\begin{document}

\begin{CJK}{UTF8}{<font>}

\title{Two-proton radioactivity within a generalized liquid drop model}
\author{J. P. Cui$^{1,2}$}
\author{Y. H. Gao$^{1,2}$}
\author{Y. Z. Wang$^{1,2,3}$}
\email{yanzhaowang09@126.com}
\author{J. Z. Gu$^{3}$}
\email{jzgu1963@ciae.ac.cn}
\affiliation{$^1$ Department of Mathematics and Physics, Shijiazhuang Tiedao University,
Shijiazhuang 050043, China\\
$^2$ Institute of Applied Physics, Shijiazhuang Tiedao University,
Shijiazhuang 050043, China\\
$^3$ China Institute of Atomic Energy, P. O. Box 275 (10), Beijing 102413,
China}
\date{\today}

\begin{abstract}
The generalized liquid drop model (GLDM) is firstly extended to study the two-proton ($2p$) radioactivity half-lives of the ground-state of nuclei.
According to the comparison between the calculated half-lives and the experimental data, it is shown that the GLDM describes the $2p$ radioactivity half-lives well.
In addition, by comparing its accuracy with other models, it is found that the GLDM has a comparable accuracy with them. Finally, the $2p$ radioactivity half-lives of some most probable candidates are predicted with the GLDM by inputting the \textit{Q}$_{2p}$ values (the released energy of the $2p$ radioactivity) from the updated AME 2016 Mass Table, which may be useful for future experiments.

\end{abstract}

\maketitle
\section{Introduction}\label{sec0}
The two-proton ($2p$) radioactivity of nuclei was predicted for the first time by Zel'dovich~\cite{Zel1960} and Goldansky~\cite{Goldansky1960,Goldansky1961} in the 1960s.
It was treated as a new exotic decay mode with the simultaneous emission of the two protons from the unbound even-Z nuclei near or beyond the proton-drip line.
In subsequent studies, the theoreticians made great efforts to predict the potential candidates of the $2p$ radioactivity.
However, the $2p$ radioactivity was not observed for a long time due to the limitation in experiment. With the development of the detection technologies and the radioactive beam facilities, the ground-state true $2p$ radioactivity (\textit{Q}$_{2p}>0$ and \textit{Q}$_{p}<0$, where the \textit{Q}$_{p}$ is the released energy of the one-proton radioactivity.)~\cite{Pfutzner2012} was firstly observed from $^{45}$Fe at GSI~\cite{Pfutzner2002} and at GANIL~\cite{Giovinazzo2002} in 2002, respectively.
Then, it was discovered in $^{54}$Zn~\cite{Blank2005}, $^{48}$Ni~\cite{Dossat2005}, $^{19}$Mg~\cite{Mukha2007}, and $^{67}$Kr~\cite{Goigoux2016}.
In fact, before 2002, the extremely short-lived ground state $2p$ radioactivity has been observed from $^{6}$Be~\cite{Whaling1966}, $^{12}$O~\cite{KeKelis1978}, and $^{16}$Ne~\cite{KeKelis1978}.
The decaying states have large widths so that the $2p$ emitter states and the $1p$ daughter states overlap, which was interpreted as the \textquotedblleft democratic decay\textquotedblright in later studies~\cite{Pfutzner2012,Bochkarev1989,Blank2008}.
The above mentioned two kinds of ground-state $2p$ emissions have become a hot subject in the field of modern nuclear physics.

In order to estimate the half-lives of the $2p$ radioactivity, many theoretical approaches have been proposed by researchers~\cite{Goldansky1960,Grigorenko2007,Galitsky1964,Lane1958,Miernik2007,Olsen2013,
Olsen2013-2,Alvarez2008,Goldansky1961,Barker2001,Brown1991,Naza1996,Grigorenko2000,
Grigorenko2002,Grigorenko2007a,Alvarez2010,Garrido2008,Vasilevsky2001,Delion2013}.
In general, these approaches can be mainly divided into two kinds.
One is the simplified theoretical models, which include the direct decay model~\cite{Grigorenko2007,Galitsky1964,Lane1958,Miernik2007,Olsen2013,Olsen2013-2}, the simultaneous versus sequential decay model~\cite{Goldansky1960,Alvarez2008}, and the diproton model~\cite{Olsen2013,Olsen2013-2,Goldansky1961,Barker2001,Brown1991,Naza1996}. For the diproton model, it is an extreme picture that the two emitted protons are correlated strongly.
So the two protons are emitted as a $^{2}$He-cluster from a parent nucleus.
The other is the three-body models~\cite{Grigorenko2000,Grigorenko2002,Grigorenko2007a,Alvarez2010,Garrido2008,Vasilevsky2001}. Among the three-body models, the model with the asymptotic three-body potential is a representative one by treating the spatial and momentum correlations in detail~\cite{Grigorenko2000,Grigorenko2002,Grigorenko2007a}.
Within these approaches, the experimental half-lives are reproduced more or less satisfactorily, but these approaches need further improvement. On the one hand, more and detailed nuclear structure information should be included in these models~\cite{Pfutzner2012,Blank2008,Rotureau2006,Grigorenko2009}. On the other hand, all different decay paths and the interference between them should be taken into account~\cite{Pfutzner2012,Blank2008,Rotureau2006,Grigorenko2009}. However, the improvement task is very difficult. Very recently, Gon\c{c}alves \textit{et al.} calculated the half-lives of the $2p$ radioactivity using the effective liquid drop model (ELDM) in the spherical nuclear approximation~\cite{Goncalves2017}.
Within the ELDM, the $2p$ emission process was treated as the $^{2}$He-cluster radioactivity and the calculated half-life values are in good agreement with the experimental data.
It is well known that the GLDM is a successful model to investigate various charged particle radioactivities~\cite{GLDM,cui2016,wang2010,wang2014,Dong2009}.
Motivated by the work of the Gon\c{c}alves \textit{et al.}, we will try extending the GLDM to the study of the $2p$ radioactivity of some proton-rich nuclei.
This article is organized as follows.
In Sec. II, the framework of the GLDM is presented. The calculated half-lives of the $2p$ radioactivity are shown and discussed in Sec. III. In the last section, some conclusions are drawn.

\section{GLDM}

In the GLDM, the process of the shape evolution from one body to two separated fragments can be described in a unified way.
Many important factors, such as the precise nuclear radius, mass and charge asymmetry, and proximate effect, are included.
It has been successfully used to calculate the half-lives of the $\alpha $-decay, the cluster radioactivity, the spontaneous fission and the one-proton emission~\cite{GLDM,cui2016,wang2010,wang2014,Dong2009}.

For the $2p$ radioactivity in the framework of GLDM, we consider it as a two-body problem.
It is assumed that a preformed $2p$ pair penetrates the Coulomb barrier of the parent nuclei and decays outside the barrier.
The $2p$ pair preformed nearby the surface of the parent nuclei has zero binding energy and the two protons separate quickly due to the dominance of the Coulomb repulsion after they escape from the parent nuclei.

In the GLDM, the macroscopic energy \textit{E}(\textit{r}) is written as
\begin{equation}  \label{etot}
E(r)=E_{V}+E_{S}+E_{C}+E_{\text{Prox}}(r)+E_{\text{cen}}(r),
\end{equation}
which contains the volume, surface, Coulomb, proximity, and centrifugal potential energies.

For one-body shapes, the volume, surface, and Coulomb energies (all in MeV) are defined
as
\begin{equation}  \label{evone}
E_{V}=-15.494(1-1.8I^2)A \ ,
\end{equation}
\begin{equation}  \label{esone}
E_{S}=17.9439(1-2.6I^2)A^{2/3}(S/4\pi R_0^2) \ ,
\end{equation}
\begin{equation}  \label{econe}
E_{C}=0.6e^2(Z^2/R_0) \times 0.5\int (V(\theta)/V_0)(R(\theta)/R_0)^3 \sin
\theta d \theta,
\end{equation}
where $A$, $Z$, $S$ and $I$ are the mass number, charge number, surface, and relative neutron excess of the parent nucleus, respectively.
$V(\theta )$ is the electrostatic potential at the surface and $V_0$ is the surface potential of
the sphere. The radius (in fm) of the parent nucleus is given by
\begin{equation}  \label{radii}
R_0=(1.28A^{1/3}-0.76+0.8A^{-1/3}) \ .
\end{equation}

When the two fragments are separated
\begin{equation}  \label{ev}
E_{V}=-15.494\left \lbrack (1-1.8I_1^2)A_1+(1-1.8I_2^2)A_2\right \rbrack \,,
\end{equation}
\begin{equation}  \label{es}
E_{S}=17.9439\left \lbrack(1-2.6I_1^2)A_1^{2/3}+(1-2.6I_2^2)A_2^{2/3} \right
\rbrack \,,
\end{equation}
\begin{equation}  \label{ec}
E_{C}=0.6e^2Z_1^2/R_1+0.6e^2Z_2^2/R_2+e^2Z_1Z_2/r,
\end{equation}
where $A_i$, $Z_i$, $R_i$, and $I_i$ are the mass numbers, charge numbers,
radii and relative neutron excesses of the two fragments.
$r$ is the distance between the two fragments.

The surface energy results from the effects of the surface tension forces in
a half space. When there are nucleons in regard in a neck or a gap between
separated fragments an additional term called proximity energy must be added
to take into account the effects of the nuclear forces between the close
surfaces. This term is essential to describe smoothly the one-body to
two-body transition and to obtain reasonable fusion barrier heights. It
moves the barrier top to an external position and strongly decreases the
pure Coulomb barrier
\begin{equation}
E_{\text{Prox}}(r)=2\gamma \int _{h_{\text{min}}} ^{h_{\text{max}}} \Phi
\left \lbrack D(r,h)/b\right \rbrack 2 \pi hdh,
\end{equation}
where $h$ is the distance varying from the neck radius or zero to the height
of the neck border. $D$ is the distance between the surfaces in regard and $%
b=0.99$~fm is the surface width. $\Phi$ is the proximity function of
Feldmeier. The surface parameter $\gamma$ is the geometric mean
between the surface parameters of the two fragments.

The centrifugal potential energy \textit{E}$_{\text{cen}}(r)$ is adopted by
the following form
\begin{equation}
E_{\text{cen}}(r)=\frac{\hbar ^{2}}{2\mu }\frac{l(l+1)}{r^{2}},
\end{equation}
where $l$ is the orbital angular momentum carried by the $2p$ pair.
$\mu$ stands for the reduced mass of the $2p$ pair and the residual daughter nucleus.

The penetrability factor $P$ is calculated by the WKB approximation, which is expressed as
\begin{equation}  \label{penetrability0}
P=\exp \left[ -\frac{2}{\hbar }\int_{R_{\text{in}}}^{R_{\text{out}}}%
\sqrt{2B(r)[E(r)-E_{sph}]}\text{d}r\right],\
\end{equation}
where \textit{R}$_{\text{in}}$ and \textit{R}$_{\text{out}}$ are the two turning points of the WKB action integral.
The two following approximations may be used: $R_{\text{in}}=R_{\text{d}}+R_{2p}$
and $B(r)=\mu$.
Because of the influence of centrifugal potential energy, $R_{\text{out}}$ is the following form
\begin{equation}  \label{rout}
R_{\text{out}}=\frac{Z_{1}Z_{2}e^{2}}{2Q_{2p}}+\sqrt{(\frac{Z_{1}Z_{2}e^{2}}{%
2Q_{2p}})^{2}+\frac{l(l+1)\hbar ^{2}}{2\mu Q_{2p}}}.
\end{equation}

The decay constant $\lambda$ is defined as
\begin{equation}  \label{decay constant1}
\lambda=S_{2p}\nu _{0}P,
\end{equation}
where $S_{2p}$ denotes the spectroscopic factor of the $2p$ radioactivity. It can be estimated in the cluster overlap approximation~\cite{Brown1991}, $S_{2p}=G^{2}[A/(A-2)]^{2n}\chi ^{2}$. Here, $G^{2}=(2n)!/[2^{2n}(n!)^{2}]$~\cite{Anyas1974}, and $n$ is the average principal proton oscillator quantum number given by $n$$\approx (3Z)^{1/3}-1$~\cite{Bohr1969}. $\chi ^{2}=0.0143$, which is determined by fitting the experimental half-lives of $^{19}$Mg, $^{45}$Fe, $^{48}$Ni, and $^{54}$Zn.

$\nu _{0}$ is the assault frequency of the $2p$ pair on the barrier of the parent nucleus and is estimated by the classical method
\begin{equation}  \label{frequency}
\mathit{\nu }_{0}=\frac{1}{2R_0}\sqrt{\frac{2E_{2p}}{M_{2p}}},
\end{equation}
where $E_{2p}$ and $M_{2p}$ represent the kinetic energy and the mass of the emitted $2p$ pair, respectively.

Finally, the $2p$ radioactivity half-life is calculated by
\begin{equation}  \label{GLDM2}
T_{1/2}=\frac{\ln 2}{\lambda}.
\end{equation}
\section{Results and Discussions}
We have performed calculations on the $2p$ radioactivity half-lives of the ground-state to ground-state transitions with the GLDM by inputting the experimental \textit{Q}$_{2p}$ values.
The calculated results are listed in TABLE I.
The first column of TABLE I denotes the parent nuclei.
In columns 2 and 3, the experimental \textit{Q}$_{2p}$ values and the experimental logarithm half-lives are shown, respectively.
The calculated logarithm half-lives by the GLDM, the ELDM~\cite{Goncalves2017}, and an empirical formula of Ref.~\cite{Sreeja2019} are listed in the 4th-6th columns.
In calculations the orbital angular momenta carried by the $2p$ pair are all zero which is determined by the spin-parity selection rule.
In order to analyze the deviation between the experimental half-lives and the calculated ones, the values of the logarithm hindrance factor $\log_{10}HF$ ($\log_{10}HF=\log_{10}T_{1/2}^{\text{exp.}}-\log_{10}T_{1/2}^{\text{GLDM}}$)
are listed in the last column of TABLE I.
From TABLE I, it can be seen that the calculated $2p$ decay half-lives by the GLDM are in agreement with the experimental data except for the case of $^{16}$Ne.
In addition, by comparing the half-lives within the ELDM~\cite{Goncalves2017} and an empirical formula of Ref.~\cite{Sreeja2019} with the experimental ones, the same conclusion can be obtained. So, it can be drawn that the GLDM has a comparable accuracy with the two other methods.

For $^{16}$Ne, the deviations between the experimental half-lives and calculated ones with the three models could be up to about 3 orders of magnitude.
Since the $2p$ decay half-lives of $^{16}$Ne were measured in 1978~\cite{KeKelis1978} and in 1983~\cite{Woodward1983} and the half-lives were as short as $10^{-21}$\,s.
So the experimental data might not be so accurate and the experimental uncertainty could be large owing to the old measurement method and detection technology. So it is suggested that the $2p$ decay half-life of $^{16}$Ne should be measured again with the current detection facilities.

Note that Mukha \textit{et al.} measured the $2p$ decay half-life of the 21$^{+}$ isomer state of $^{94}$Ag~\cite{Mukha2006}. The measured half-life is 80\,s. However, the calculated half-life with the GLDM is as long as 5.0$\times 10^{4}$\,s. Previous studies suggested that the 21$^{+}$ isomer of $^{94}$Ag is assumed to be very strongly deformed  nucleus~\cite{Pfutzner2012,Goncalves2017,Sreeja2019,Mukha2006}. A recent work of Goigoux \textit{et al.} suggested that the deformation of a nucleus played crucial role in the $2p$ decay half-life~\cite{Goigoux2016}. However, in the GLDM and ELDM, the deformation of the ground state is not taken into account. For the empirical formula of Ref.~\cite{Sreeja2019}, the parameters are fitted by the predicted $2p$ decay half-lives of the ELDM. This may be the reason why the experimental half-lives of the 21$^{+}$ isomer of $^{94}$Ag are not reproduced by the three models. In addition, some researchers pointed out that the three-body asymptotic behavior and configuration mixing are the key factors for the $2p$ radioactivity~\cite{Grigorenko2009,Goigoux2016,Rotureau2006}.
Therefore, it is important to improve the GLDM and ELDM by considering the nuclear deformation, three-body asymptotic behavior and configuration mixing to make reliable predictions for the $2p$ decay half-lives in the future. Nevertheless, as a whole the GLDM is a successful model to study the $2p$ decay half-lives of the ground-state nuclei.
\linespread{1.5}
\begin{table*}[t]
\begin{ruledtabular}\caption{The experimental and calculated $2p$ radioactivity half-lives of the ground state to ground state transitions. The \textit{Q}$_{2p}^{\text{exp.}}$ and log$_{10}T_{1/2}$ values are measured in MeV and seconds, respectively.}

\begin{tabular}{llllccc}
Nuclei & \textit{Q}$_{2p}^{\text{exp.}}$(MeV)   & log$_{10}T_{1/2}^{\text{exp.}}$(s) &
log$_{10}T_{1/2}^{\text{GLDM}}$(s) &log$_{10}T_{1/2}^{\text{ELDM}}$(s)~\cite{Goncalves2017} &log$_{10}T_{1/2}^{\text{for.}}$(s)~\cite{Sreeja2019}&log$_{10}HF$ \\
\hline

$^{6}_{4}$Be&1.371(5)~\cite{Whaling1966}  &-20.30$_{-0.03}^{+0.03}$~\cite{Whaling1966} &-19.37$_{-0.01}^{+0.01}$ &-19.97 &-21.95 &-0.94\\
$^{12}_{8}$O&1.638(24)~\cite{Jager2012}  &$>$-20.20~\cite{Jager2012} &-19.17$_{-0.08}^{+0.13}$ &-18.27 &-18.47&$>$-1.02 \\
        &1.820(120)~\cite{KeKelis1978}  &-20.94$_{-0.21}^{+0.43}$~\cite{KeKelis1978} &-19.46$_{-0.07}^{+0.13}$ &-&- &-1.48 \\
        &1.790(40)~\cite{Kryger1995}  &-21.10$_{-0.13}^{+0.18}$~\cite{Kryger1995} &-19.43$_{-0.03}^{+0.04}$ &-&- &-1.67 \\
        &1.800(400)~\cite{Suzuki2009}  &-21.12$_{-0.26}^{+0.78}$~\cite{Suzuki2009} &-19.44$_{-0.20}^{+0.30}$ &-&- &-1.68 \\
$^{16}_{10}$Ne&1.330(80)~\cite{KeKelis1978}  &-20.64$_{-0.18}^{+0.30}$~\cite{KeKelis1978} &-16.45$_{-0.21}^{+0.23}$ &-&-15.94&-4.20 \\
         &1.400(20)~\cite{Woodward1983}  &-20.38$_{-0.13}^{+0.20}$~\cite{Woodward1983} &-16.63$_{-0.05}^{+0.05}$ &-16.60 &-16.16 &-3.75\\
$^{19}_{12}$Mg&0.750(50)~\cite{Mukha2007}  &-11.40$_{-0.20}^{+0.14}$~\cite{Mukha2007} &-11.79$_{-0.42}^{+0.47}$ &-11.72 &-10.66&0.40 \\
$^{45}_{26}$Fe&1.100(100)~\cite{Pfutzner2002}  &-2.40$_{-0.26}^{+0.26}$~\cite{Pfutzner2002} &-2.23$_{-1.17}^{+1.34}$ &-&-&-0.17\\
         &1.140(50)~\cite{Giovinazzo2002}  &-2.07$_{-0.21}^{+0.24}$~\cite{Giovinazzo2002} &-2.71$_{-0.57}^{+0.61}$ &-&-1.66&0.64 \\
         &1.154(16)~\cite{Dossat2005}  &-2.55$_{-0.12}^{+0.13}$~\cite{Dossat2005} &-2.87$_{-0.18}^{+0.19}$ &-2.43 &-1.81&0.32 \\
         &1.210(50)~\cite{Audirac2012}  &-2.42$_{-0.03}^{+0.03}$~\cite{Audirac2012} &-3.50$_{-0.52}^{+0.56}$ &-&-2.34&1.08 \\
$^{48}_{28}$Ni&1.290(40)~\cite{Pomorski2014}  &-2.52$_{-0.22}^{+0.24}$~\cite{Pomorski2014} &-2.62$_{-0.42}^{+0.44}$ &-&-1.61 &0.10\\
         &1.350(20)~\cite{Dossat2005}  &-2.08$_{-0.78}^{+0.40}$~\cite{Dossat2005} &-3.24$_{-0.20}^{+0.20}$ &-&-2.13 &1.16\\
          &1.310(40)~\cite{Wang2017}  &-2.52$_{-0.22}^{+0.24}$~\cite{Pomorski2011} &-2.83$_{-0.41}^{+0.43}$ &-2.36&- &0.31\\
$^{54}_{30}$Zn&1.280(210)~\cite{Ascher2011}  &-2.76$_{-0.14}^{+0.15}$~\cite{Ascher2011} &-0.87$_{-0.24}^{+0.25}$ &-&-0.10&-1.89 \\
         &1.480(20)~\cite{Blank2005}  &-2.43$_{-0.14}^{+0.20}$~\cite{Blank2005} &-2.95$_{-0.19}^{+0.19}$ &-2.52 &-1.83&0.51 \\
$^{67}_{36}$Kr&1.690(17)~\cite{Goigoux2016}  &-1.70$_{-0.02}^{+0.02}$~\cite{Goigoux2016} &-1.25$_{-0.16}^{+0.16}$ &-0.06 &0.31&-0.45\\
\end{tabular}
\end{ruledtabular}
\end{table*}

Encouraged by the above discussions, we will attempt to predict the half-lives of the most probable $2p$ decay candidates with the GLDM.
Since the decay energy plays a crucial role in charged particle radioactivities~\cite{yzwang2015,cui2018}, the \textit{Q}$_{2p}$ and \textit{Q}$_{p}$ values in the calculation are extracted from the updated AME 2016 Mass Table~\cite{Wang2017}, they are listed in columns 2 and 3 of TABLE II.
Since the ground-state $2p$ radioactivity is usually found in the even-Z nuclei due to the proton-proton paring effect, we only select the even-Z candidates of \textit{Q}$_{2p}>0$ from Ref.~\cite{Wang2017}. In calculations, the orbital angular momenta are all used as zero by the spin-parity selection rule. By using the GLDM, the predicted $2p$ decay half-lives of the light and medium mass nuclei are presented in the last column of TABLE II.
The nuclei in TABLE II are divided into the following two categories: the nuclei of the true ${2p}$ radioactivity (\textit{Q}$_{2p}>0$, \textit{Q}$_{p}<0$) and the ones of not the true ${2p}$ radioactivity (\textit{Q}$_{2p}>0$, \textit{Q}$_{p}>0$).
For the case of the true ${2p}$ radioactivity, $^{22}_{14}$Si, $^{34}_{20}$Ca, $^{39}_{22}$Ti, and $^{42}_{24}$Cr should be paid attention to by experimental researchers, because the predicted half-lives of the four candidates are located in the range of the measured $2p$ radioactivity half-lives. However, for the other four $2p$ candidates ($^{49}_{28}$Ni, $^{55}_{30}$Zn, $^{60}_{32}$Ge, and $^{64}_{34}$Se), their half-lives are very long, which would be difficult to be observed in measurement.

For the case of the not true ${2p}$ radioactivity, $^{26}_{16}$S, $^{38}_{22}$Ti, $^{58}_{32}$Ge, and $^{59}_{32}$Ge should also attract the attention of researchers because each of their \textit{Q}$_{2p}$ values is much larger than the corresponding \textit{Q}$_{p}$ value. Thus, the ${2p}$ radioactivity may be the dominant decay mode for them.

\linespread{1.3}
\begin{table*}[t]
\begin{ruledtabular}\caption{The predicted $2p$ decay half-lives (in seconds) of the proton-rich nuclei with the GLDM by inputting the \textit{Q}$_{2p}$ values (in MeV) extracted from Ref.~\cite{Wang2017}. The \textit{Q}$_{p}$ values are also measured in MeV. }
\begin{tabular}{lllll}
Nuclei &\textit{Q}$_{2p}$(MeV) &\textit{Q}$_{p}$(MeV)&$T_{1/2}^{\text{GLDM}}$(s) \\
\hline
&&True $2p$ radioactivity\\
\hline
$^{22}_{14}$Si&1.28&-0.94 &5.06$\times 10^{-14}$\\
$^{34}_{20}$Ca&1.47&-0.48 &1.95$\times 10^{-11}$\\
$^{39}_{22}$Ti&0.76 &-0.84&4.56$\times 10^{-2}$\\
$^{42}_{24}$Cr&1.00 &-0.88&1.33$\times 10^{-3}$\\
$^{49}_{28}$Ni&0.49&-0.59 &2.90$\times 10^{14}$\\
$^{55}_{30}$Zn&0.48 &-0.45&8.73$\times 10^{17}$\\
$^{60}_{32}$Ge&0.63&-0.62 &3.52$\times 10^{13}$\\
$^{64}_{34}$Se&0.46 &-0.49&2.76$\times 10^{24}$\\
\hline
&& Not true $2p$ radioactivity\\
\hline
$^{26}_{16}$S&1.76 &0.05&2.55$\times 10^{-15}$\\
$^{38}_{22}$Ti&2.74 &0.06 &5.36$\times 10^{-15}$\\
$^{58}_{32}$Ge&3.73&0.64  &8.02$\times 10^{-14}$\\
$^{59}_{32}$Ge&2.10 &0.38 &1.06$\times 10^{-7}$\\

\end{tabular}
\end{ruledtabular}
\end{table*}

\section{Conclusions}
In this article, the $2p$-decay half-lives of the known proton-rich nuclei have been calculated with the GLDM. According to the comparison between the experimental half-lives and the calculated ones, it is found that the experimental half-lives can be reproduced well within the GLDM. Then, the GLDM has been used to predict the most probable $2p$-decay half-lives of the light and medium mass nuclei by inputting the \textit{Q}$_{2p}$ values from the updated AME 2016 Mass Table. By analyzing the predicted results, the candidates of $^{22}$Si, $^{34}$Ca, $^{39}$Ti, and $^{42}$Cr are suggested and they should be paid attention to by experimental researchers because the four nuclei prefer to the true $2p$ radioactivity. We hope this prediction is helpful for future measurements.

\section*{ACKNOWLEDGEMENTS}
We thank professors Guy Royer, Shangui Zhou, Fengshou Zhang and Ning Wang for their helpful discussions. This work was supported by the National Natural Science Foundation of China (Grant No. U1832120 and No. 11675265) and the Natural Science Foundation for outstanding Young Scholars of Hebei Province of China (Grant No. A2018210146). \label{sec4}

\end{CJK}
\end{document}